# Turing's cascade instability supports the coordination of the mind, brain, and behavior

**Damian G. Kelty-Stephen[1] and Madhur Mangalam[2]**

[1]*Department of Psychology, State University of New York at New Paltz, New Paltz, NY, USA*

[2]*Department of Physical Therapy, Movement and Rehabilitation Sciences, Northeastern University, Boston, MA, USA*

ORCIDs:

0000-0001-7332-8486 (D. G. Kelty-Stephen)

0000-0001-6369-0414 (M. Mangalam)

Correspondence:

keltystd@newpaltz.edu (D. G. Kelty-Stephen)

m.mangalam@northeastern.edu (M. Mangalam)

## Abstract

Turing inspired a computer metaphor of the mind and brain that has been handy and has spawned decades of empirical investigation, but he did much more and offered behavioral and cognitive sciences another metaphor—that of the cascade. The time has come to confront Turing's cascading instability, which suggests a geometrical framework driven by power laws and can be studied using multifractal formalism and multiscale probability density function analysis. Here, we review a rapidly growing body of scientific investigations revealing signatures of cascade instability and their consequences for a perceiving, acting, and thinking organism. We review work related to executive functioning (planning to act), postural control (bodily poise for turning plans into action), and effortful perception (action to gather information in a single modality and action to blend multimodal information). We also review findings on neuronal avalanches in the brain, specifically about neural participation in body-wide cascades. Turing's cascade instability blends the mind, brain, and behavior across space and time scales and provides an alternative to the dominant computer metaphor.

# Glossary

**Akaike weight:** Model averaging can be done with Akaike weights that represent a model's relative likelihood. First the model's relative likelihood is calculated, which is simply exp(-0.5 * the AIC score for that model). This value is divided by the sum of these values across all models to get the Akaike weight for a model.

**Affordance:** An affordance is what an organism can do with an object or a surface based on its capabilities.

**Autocorrelated:** A measured time series that shows correlation between its elements separated by a given interval.

**Avalanches:** A series of bursts of activity [in neural networks] that can be described by a power law in terms of size distribution.

**Cascade:** A physical process characterized by a blending of information or structure built at multiple scales. For instance, when events spread from one scale to another, e.g., cellular to genetic, or cellular to whole-tissue, and then to organ- and to whole-organism scale, what we have is a cascade of effects.

**Criticality:** Criticality is a phenomena that produces power-law distributed avalanche sizes in certain complex systems with several interacting components, such as neural networks, forest fires, and power grids.

**Critical branching:** A branching process is a form of stochastic process that consists of collections of random variables. This processes is considered to show critical branching when it produces power-law distribution of different-sized events in the study system.

**Drunkard's walk:** A method for calculating the likely location of a point subject to random motions, given the odds (which are the same at each step) of moving some distance in some direction.

**Dynamic or effortful touch:** The act of perceiving physical properties such as heaviness and length of an object held by the hand or attached the body by means of muscular effort, that is, wielding.

**Intermittency:** An uneven distribution of events across time or space, characterized by the unpredictable, sudden appearance or disappearance of stability—or by the unpredictable, sudden transitions among stable states.

**Log-likelihood:** A measure of how well a particular model fit the data.

**Monofractality:** A statistical index of complexity describing how detail in a fractal pattern changes with the scale at which it is measured.

**Multifractality:** A generalization of a fractal system in which one fractal dimension is not enough to describe its dynamics; instead, a continuous spectrum of exponents (the so-called singularity spectrum) is needed.

**Neural tuning: The ability of a neuron** to selectively reflect one type of sensory, associative, motor, or cognitive information. This ability can change with experience, and the neuron is considered to have changed its tuning.

**Nonstationary:** A nonstationary process that has an unconditional joint probability distribution that changes with time. As a result, parameters like mean and variance also fluctuate with time.

**Phase:** A discrete time or stage in a sequence of events or a change or development process, usually associated with the aplitude of a waveform.

**Poisson-like processes:** A Poisson-like process, named after French mathematician Siméon Denis Poisson, entails discrete probability distribution of a given number of events occurring in a fixed interval of time or space with a known constant mean rate and independent of the time since the last event.

**Postural center of pressure (CoP):** The resultant vertical force vector on the supporting surface where all force vectors could be considered to have a single point of application. A shift in the CoP is an indirect measure of postural sway and consequently a measure of one's ability to maintain balance.

**Postural sway:** The subtle, unconscious movements that occur around the center of gravity of the body in order to maintain balance. It is the body's normal response to changing environmental conditions. It could be obvious or undetectable.

**Power law:** A power law is a functional relationship between two quantities in which a change in one quantity causes a corresponding change in the other quantity, regardless of their original sizes: one quantity changes as a power of another.

**Rambling:** In quiet standing, the COP displacement can be decomposed into two componets. The reference point migration is called rambling.

**Shape collapse:** The observation that long-duration avalanche profiles have the same scaled mean shape as short-duration avalanches, which can be captured by a universal scaling function.

**Susceptibility:** The likelihood that a neuronal avalanche will occur in a region based on the characteristics and conditions of local network connectivity.

**Tensegrity:** The property of a stable three-dimensional structure consisting of contiguous members under tension and non-contiguous members under compression.

**Trail Making Test (TMT):** A clinical test of the executive function of rule switching that requires tracing trajectories between letters and numbers across a spatial display.

**Trembling:** In quiet standing, the COP displacement can be decomposed into two componets. The reference point migration is called rambling and the COP migration around the reference is coined trembling.

# 1. Introduction

The present work is the story of a metaphor long overshadowed by its twin. It has grown and has spread, and though it has not changed, we have learned more about it and how to make scientific use of it. This long under-appreciated metaphor has come home to the behavioral and cognitive sciences, and it may soon unseat its more dominant metaphor or at least require us, behavioral and cognitive scientists, to reckon with our commitments to each of these twin metaphors. We refer to the twin metaphors given to us by Turing: the computer metaphor (1950) and the cascade-instability metaphor (Kelso, 1995; Turing, 1952; see also Hodges, 1983). The computer metaphor is most dominant in behavioral and cognitive sciences (Atkinson and Shiffrin, 1968; Boden, 1988; Churchland and Sejnowski, 1994; Pinker, 2021; Simon, 1969; Von Neumann, 1958; Wolpert and Ghahramani, 2000). The cascade metaphor has sooner taken hold in the biological sciences, slipping into the cracks between genetic determinism and phenotypic (Turing, 1952) and spreading through the more fluid aspects of biological form (Neubert et al., 2002; Rätz and Röger, 2012; Woolley et al., 2021). Turing (1952) warned that the full entailments of the cascade metaphor were beyond the frontiers of current mathematical modeling, and pushing on that frontier was some of his last work.

   To date, cascade-inspired modeling has blossomed, and it might be that the cascade instability is ready now to step out of the dark shadow casted by the computer metaphor. Cascade modeling may offer a new theoretical foothold in explaining how the fluidity of organism behavior can support perception, action, and cognition—in a way that we had typically reserved for the computer metaphor. The notion of cascade instability is not new. Turing (1950) coined the idea himself, but he was also picking up on ideas that had been floating around in the scholarly winds (James, 1892; Richardson, 1930), and it has reappeared repeatedly since then (Froese et al., 2013; Thelen and Smith, 2006). This cascade instability has been a long-struggling metaphor for the mind struggling to assert itself on equal footing as the computer. What we think has brought the matter to a tipping point is the elaboration of multifractal modeling. Multifractal modeling provides a reliable and versatile empirical anchoring of cascade instability (Ihlen and Vereijken, 2010; Lovejoy and Schertzer, 2018; Mandelbrot, 1974). And we can find multifractal geometries spreading throughout the human organism, offering a common substrate for the context-sensitive coordination among the mind, brain, and behavior. The present review lays out the depth and extent of multifractal evidence for how cascade instability might coordinate an organism's perception, action, and cognition.

# 2. Turing's cascade instability

Alan Turing's last scholarly efforts addressed the intriguing proposal that seemingly complex patterns in nature might arise from random fluctuations (Turing, 1952; Fig. 1). At first glance, this proposal may sound absurd—the prevailing statistical pedagogy in undergraduate and graduate classes leaves most of us rooted in the general linear model. At the heart of the linear model is the theorem that many independent random events collapse towards a Normal distribution centered on an average and spread out evenly in all directions. Karl Pearson called this progression a '**drunkard's walk**' because the many independent random events entailed an incoherent progression towards ultimate stasis: random events ending more or less where they began (Pearson, 1905). So, it may strike us as odd that random fluctuations should give us patterns—or anything except more randomness.

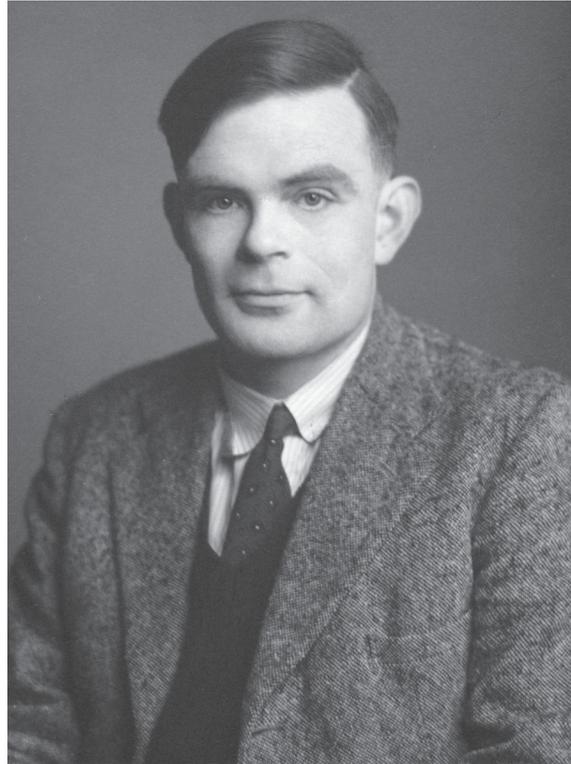

**Fig. 1.** Alan Turing (1912– 1954). Copyright © The Royal Society.

Turing's idea begins to make much more sense when we consider that the oddity sooner belongs to the notion of independence, not of order-from-noise. The drunkard's random walk presumes independence among random fluctuations and this presumption is a wonderful logical step to produce a reliable geometrical model of random variation. However, independence is the limiting case—valuable for the textbook demonstration but rarely available in practice. Random fluctuations rarely occur in isolation, and many fluctuations in the same neighborhood will sooner collide and interact than avoid each other. Whereas independent random fluctuations will wipe out any progression or development, interdependent, interacting fluctuations will propel themselves to new configurations (Fig. 2). Rather than reinventing existing heterogeneities like a drunkard's walk, the interdependent random fluctuations will use existing heterogeneities as a springboard to ever-new configurations. Interdependent fluctuations can engender interactions across many spatial and temporal scales, reshaping the context for future fluctuations. For instance, cascade instabilities can beget turbulent structures, complex flows in which once-parallel currents collapse or explode into a dizzying, perhaps uncountable diversity of vortices and eddies, with intermittent swellings and ebbings across space and time (Mandelbrot, 1974; Shlesinger et al., 1987).

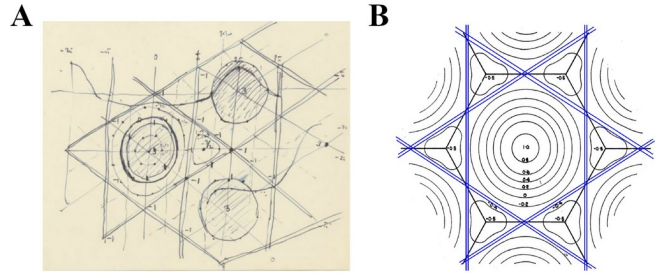

**Fig. 2.** Hexagonal planforms for patterns. (**A**) The original sketch by Turing. Reproduced from AMT/C/27/19a. Copyright © W.R. Owens. (**B**) Reproduced sketches from Pellew and Southwell (1940). The pairs of blue thick solid lines have been added to relate the two figures.

Turing saw instability not only as the root of biological pattern formation (Fig. 3) such as "rosette" spots of a jaguar but as a model for creative, intelligent thought. Some portrayals emphasize Turing's interest in pattern formation as a later, separate scholarly thread from his earlier business of developing logical machinery to mimic intelligence (Dawes, 2016; Sprevak, 2017). However, while imagining various protests against the idea of mechanical intelligence, Turing offered that he saw no qualitative difference between a creative intelligence capable of sudden, swift discovery or insight and a nuclear-fission pile going critical (Turing, 1950). Even while acknowledging the value of logical inference for modeling the mind, Turing was open to the idea that physical instability could play no small complementary role.

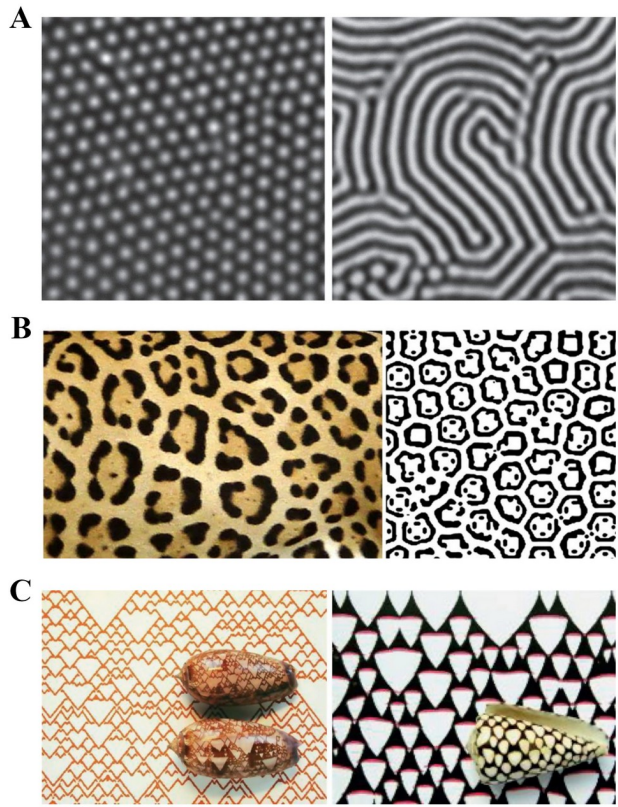

**Fig. 3.** Turing's cascade instability can lead to complex patterns observed in the nature. (**A**) Illustrative generic patterns of an activator–inhibitor scheme. From Ball (2015), courtesy of Jacques Boissonade and Patrick De Kepper, University of Bordeaux, France. (**B**) Naturally-observed "rosette" spots of a jaguar and its analog produced by two coupled activator–inhibitor processes. From Liu et al. (2006), © American Physical Society. (**C**) Naturally-observed patterns on seashells and their analogues produced by theoretical activator–inhibitor systems. From Meinhardt (2009), courtesy of Hans Meinhardt, MPI for Developmental Biology, Tübingen, Germany.

Fast forward several decades, and we find a portrait of the brain and the cognitive architecture that confirms Turing's vision of instability-driven patterns. The interdependent propagation of fluctuations in nonlinear dynamical patterns is now a commonplace fact of brain activity (Ghosh et al., 2019; Werner, 2010). The brain is understood to be a 'critical' system or show **criticality** characterized by random fluctuations interacting across many scales at once. Instead of a nuclear fissioning pile, we have the cortical columns generating spreading activity reaching from neuron to neuron and across global resting-state networks. This **cascade** instability is apparent in all aspects of the brain and behavior.

## 3. Data-analytic strategies to identify signatures of cascade instability

Multiple data-analytic strategies can help uncover or quantify cascade structure in biological and psychological measurements (Fig. 4). The preponderance of these strategies involves estimating the strength of **power-law** scaling, the tendency for measurements to show events growing or decreasing according to a power-law function of measurement scale. In large part, there are two classes of such analyses. The first involves examining a given measurement, such as a measurement series across space or time to assess how it varies across many measurement scales. Essentially, this first class of methods applies a set of grids meant to tile the measurement into nonoverlapping bins; each grid comprises bins of uniform size, and each new grid applied to the measurement contains progressively larger bins. So, one can estimate, say, the proportion or the standard deviation of the measurement in each grid's bin, and we can plot the bin statistic as a function of bin size to examine whether these statistics scale as a power-law across different bin sizes. The second class of cascade-assessing analyses involves testing for power-law scaling in histograms or probability-distribution functions (PDFs). In this second class, a measurement is examined by counting individual events (e.g., **avalanches**) within the measure, and the events of each size (or range of sizes) are counted. Cascade instability will often entail that the frequency of an event is an inverse power law of event size; that is, larger events are progressively rarer, with the ratio of dwindling frequency being proportional to the ratio of increasing size.

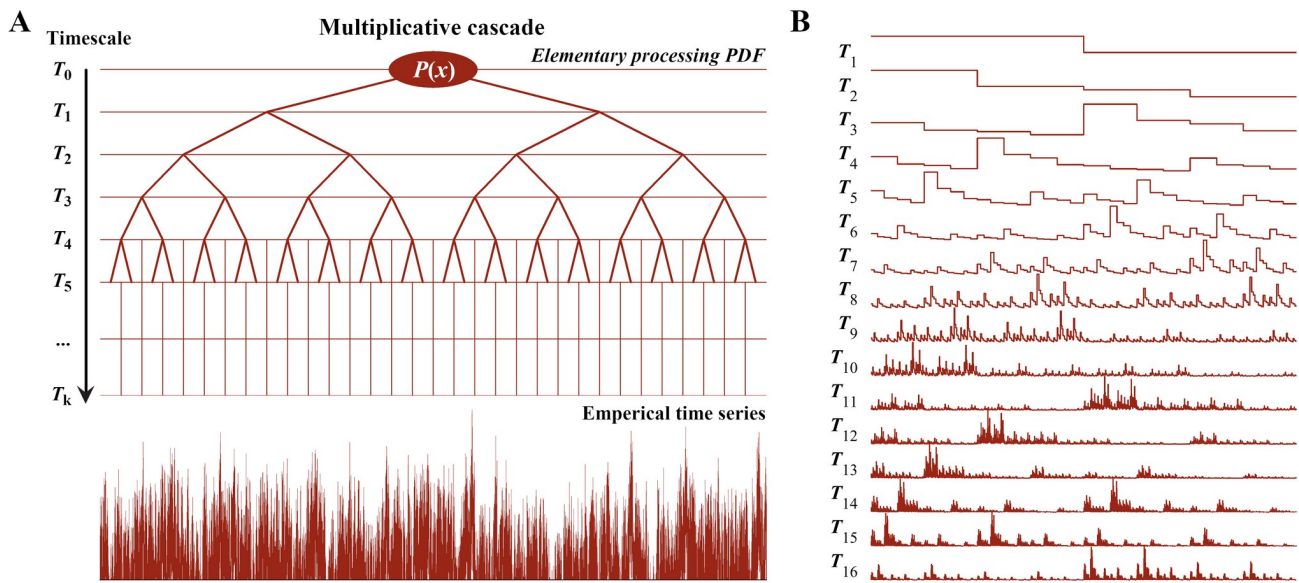

**Fig. 4.** Multiplicative cascade process. (A) Schematic of a multiplicative cascade process at increasingly fine binary timescales, driven by a general pdf, $P(x)$, starting with the whole time interval at $T_0$ and increasing the number of intervals by an integer power of two until the number of intervals at time scale $T_k$ is $2^k$. The activity time series at the bottom of the graph is believed to be formed by a multiplicative cascade, with $P(x)$ being a Gaussian density with mean $\mu$ and variance $\sigma^2$. (B) A binomial multiplicative cascade as a mathematical concept of how proportions of events can distribute themselves across progressively smaller sample sizes. The bottom-most curve shows the $2^{16}$-samples series representing the cascade after 16 generations. Each value reflects the multiplication of different sequences of the proportions 0.25 and 0.75. The leftmost samples reflect the successive multiplications of predominantly smaller proportions, and the rightmost samples reflect the successive multiplications of predominantly larger proportions. Each curve is normalized by its maximum value and displaced by 1 unit in the vertical for clarity.

Demonstrating the power-law form is often just the first step of empirically assessing cascade instability. Indeed, a cascade process will generate a power-law form in the preceding types of analysis. However, finding a power-law relationship is only consistent with linear structure and not conclusive of cascade instability, as some linear models can as well generate power-law scaling (Shlesinger et al., 1987; Wagenmakers et al., 2004). What is needed for both grid-based and PDF-based analyses of cascade instability is twofold: first, elaboration of the scaling form that each analysis estimates—we detail these elaborations in the following paragraphs; second, a way to compare the power-law results for the original data with results for the same procedure on **phase**-randomized surrogates—that is, synthetic data that preserves the linear structure of the original measurement series but contains none of the original sequences. Cascade instability entails a progressive interaction across timescales. Diagnosing cascade instability needs to ensure that the power-law form reflects aspects of the original measurement series and not the series' average linear structure irrespective of sequence.

The elaboration of the grid-based methodology of evaluating evidence of cascades is called **multifractal** analysis. The exact method of multifractal analysis can be found detailed in Chhabra & Jensen (1989) and Kantelhardt et al. (2002), who originally established these methods, as well as in tutorials by Ihlen (2012) and Kelty-Stephen et al. (2022, 2013). To summarize what multifractal analysis does: multifractal analysis examines the heterogeneity of a measurement series by modeling the scaling relationship according to which the proportion of total area under the curve increases with timescale. It estimates this scaling relationship for differently-sized events in the same measurement series and the variability of estimable scaling relationships (i.e., the "multifractal spectrum width") to quantify heterogeneity. It is possible to test for and quantify multifractal nonlinearity by calculating a *t*-statistic that compares the original measurement series' multifractal spectrum width to the multifractal spectrum width for phase-randomized surrogate series that use the same values and same autocorrelation function as in the measurement series but sorting the values in a different order to destroy the original sequence (Fig. 5).

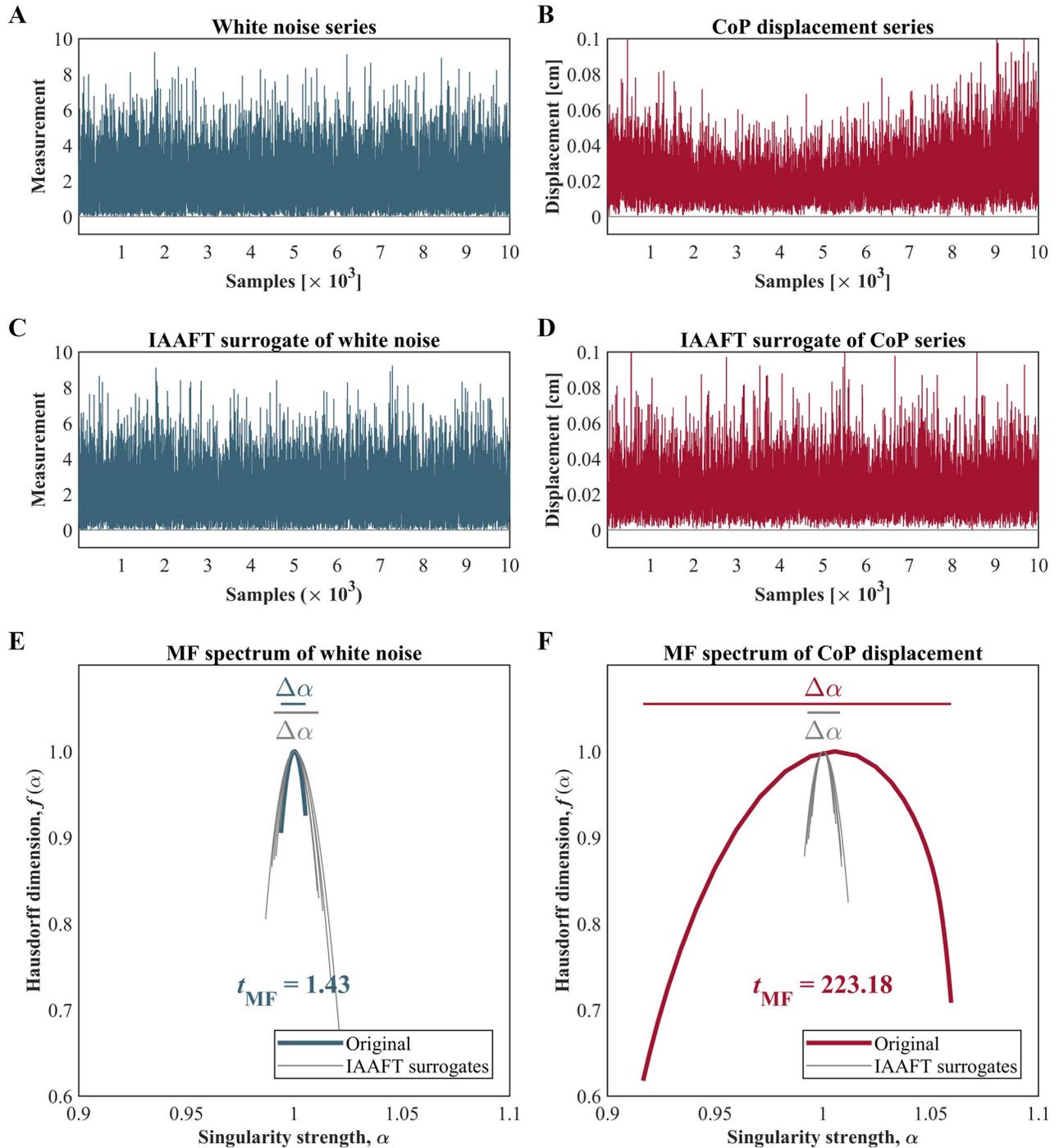

**Fig. 5.** Multifractal analysis used to quantify the strength of multifractal nonlinearity. (**A**) Simulated white noise series. (**B**) Empirically obtained postural center of pressure (CoP) displacement series of a person standing quietly with the eyes fixated at a distant point [data from Mangalam et al. (2021)]. (**C**) IAAFT surrogate of white noise in (**A**). (**D**) IAAFT surrogate of CoP displacement series in (**B**). (**E**) No significant difference in multifractal spectrum width of white noise series and multifractal spectra widths of its IAAFT surrogates ($N = 32$) suggests an absence of multifractal nonlinearity. Only five IAAFT spectra are shown for clarity. (**F**) Significantly larger multifractal spectra width of CoP displacement series than multifractal spectrum widths of its IAAFT

surrogates ($N = 32$) suggests multifractal nonlinearity. Spectra for only five of the 32 IAAFT surrogate series are shown for clarity.

The elaboration of the histogram- or PDF-based methodology involves examining how the PDF changes as the measurement for events is examined at many different timescales (Fig. 6). So, in a sense, just like the cascade-seeking analyses do end up needing to evaluate multiple curves, this time, the curve of a heavy-tailed PDF. This second class of analyses runs on power-law frequency relationships with event size but, contrary to traditional analyses, these analyses rely on empirical estimates of PDF. Maximum likelihood estimation of PDF shape is traditionally performed by estimating the **Akaike weights** for maximum likelihood estimates for exponential and lognormal distributions. The distribution chosen for a given set of measurements is determined by which distribution has the higher Akaike weight—which is itself determined by the **log-likelihood** of the distribution function and the number of parameters needed to specify that function. Power-law PDFs distinguish themselves with long, heavy tails. However, two major constraints make the proper interpretation of these tails difficult: first, tails are underpopulated, so estimates of tail heaviness are prone to small-sample bias. There is an alternate strategy called multiscale PDF analysis to address the possibility that power-law scaled PDFs could arise from **autocorrelated**, **nonstationary**, but non-cascade processes (Kiyono et al., 2007). To summarize what this analysis does: multiscale PDF analysis assesses the excess of lognormal variance beyond Gaussian variance, and it does that for an integrated series after detrending the integrated series over progressively longer bins. The same procedure is applied to a phase-randomized surrogate of the same measurement series for surrogate testing. A steep reduction in non-Gaussianity with increasing timescales in the original measurement series but a rather flat curve for the surrogates is interpreted as evidence of multiplicative cascades; that is, cascade processes had generated the measurement series.

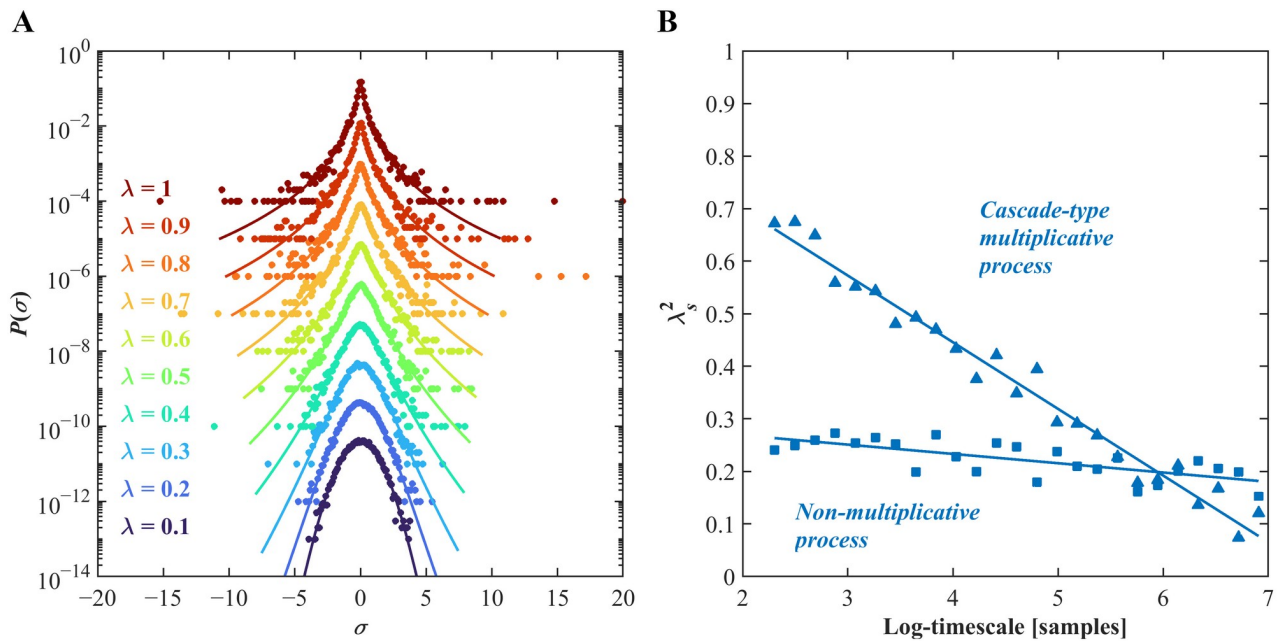

**Fig. 6.** Multiscale probability density function analysis for characterizing cascade-type multiplicative processes by estimating an index of non-Gaussianity, $\lambda$, in time series. **(A)** PDFs of numericlly simulated stochastic time series with $\lambda$ = 0.1 to 1. As $\lambda_s^2$ increases, the PDF becomes increasingly peaked and fat-tailed (dark red). As $\lambda_s$ decreases, the PDF increasingly resembles the Gaussian (dark blue), assuming a perfect Gaussian as $\lambda_s^2$ approaches 0. The PDFs have been shifted vertically for convenience of presentation, and hence, the vertical axis is given in arbitrary units. **(B)** Cascade-type multiplicative processes yield the inverse relationship $\lambda_s^2 \sim -\log s$ ($s$ = timescale, here in terms of samples). See Kiyono et al. (2007) for mathematical details of the analysis and Mangalam and Kelty-Stephen (2021) for an illustrative application in behavior and cognitive science.

## 4. Cascade instability in executive functioning: Planning to act

Psychologists and neuroscientists are beginning to consider the possibility that the explanation of executive functioning lies in generic principles of complexity. This possibility marks a substantial departure from a long history of psychometric attempts to validate models of executive control as an amalgam of many independent components. This perspective on executive function invokes cascade-driven multiplicative interactions across timescales of activity instead of isolated cognitive mechanisms. Support for this hypothesis comes from findings that multiplicative cascades mediate perceptuomotor performance in tasks involving executive control, withthe most prominent being the self-organization of a novel cognitive structure in the cognitive task of solving gear-system problems (Stephen et al., 2009a, 2009b; also see Stephen and Dixon, 2008). Likewise, Stephen and Dixon (2011) have demonstrated the role of multiplicative cascades in synchronizing behavior with an unpredictable signal; synchronizing with uncertainty remains a faculty beyond the explanatory scope of traditional models.

Multiplicative interactions in the perception-action system could support 'criticality' as a flexible poise for executive function. In this sense, these complexity-themed approaches see the same potential that Turing did. A broad agreement exists that criticality is a useful way to describe how physical models could manifest the flexibility needed for the executive function to be adaptive and context-sensitive. However, criticality as a transitional mode between multiple phases differs from criticality as the prediction of a specific computational model called 'self-organized criticality' (SOC). SOC is a relatively constrained framework that only recognizes two timescales of activity and interactivity over local spatial connectivities. The rhetorical value of SOC versus the compatibility of empirical facts with alternative multiplicative-cascade routes to criticality has been a matter of lively debate. For instance, on the one hand, so long as SOC provides a 'self-organizing' route to criticality, it may not matter whether SOC is the most accurate, exhaustive portrayal of executive function. However, on the other hand, Stephen et al. (2012) have suggested that multiplicative cascades provide a more realistic model that can produce many of the same results that SOC originally sought to

explain—that is, multiplicative cascades offer a broader class of routes for critical self-organization (Fig. 7). For instance, SOC can be situated in the broader set of cascade models—whereas SOC was originally developed to explain a single fractal (i.e., 'monofractal') pattern, it is a limiting case of more general cascade models that reliably produce multifractal results. So, SOC approaches to executive function were an important first step towards acknowledging the broader cascade framework supporting executive functions. The criticality crucial to executive functions may sooner depend on multifractal scaling than **monofractal** scaling (Kelty-Stephen and Wallot, 2017). In a bold attempt, Dixon et al. (2012) have thoroughly expanded the scope of these findings and highlighted how multifractal approaches can help understand atypical developmental outcomes and predict cognitive change.

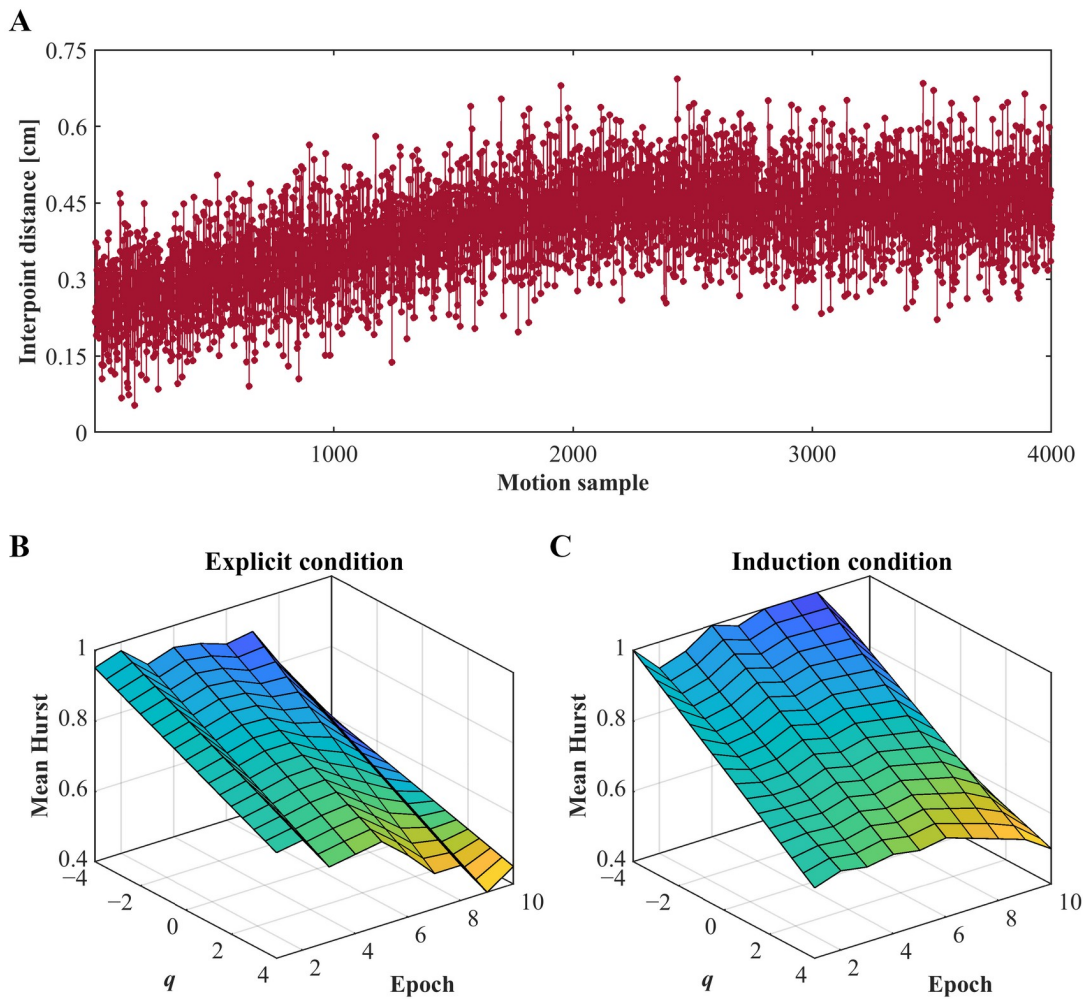

**Fig. 7.** Multifractality in card-sorting behaviors according to a rule. Participants had to sort cards using their hands, one at a time, into groups defined by a rule. Experimenters randomly assigned participants to one of two groups, one of whom had to induce the rule from feedback on individual card sortings, and the other received explicit instruction on the rule from the experimenter at the beginning of the task. All participants' sorting-hand movements were measured using a magnetic motion-capture system to calculate the Euclidean displacement series between consecutive positions for subsequent analysis of the average multifractal trajectory across time in the task in each condition. (**A**) A sample interpoint distance time series, for the induction condition. The calculated Euclidean distance between each point is plotted against motion-capture sample. (**B**) The mean $H(q)$ curve by condition plotted across $q$ as well as over epochs leading up to rule mastery. As can be seen, the trajectory of $H(2)$ is simply decreasing in the explicit condition whereas $H(2)$ follows more of an arc in the induction condition, first increasing and then decreasing on the approach to rule mastery. We sought to use vector error-correction (VEC) modeling to test whether $H(q)$ for other $q$ contributed to later changes in $H(2)$. See Stephen et al. (2012) for further details.

Computational modeling efforts, such as those by Kello (2013), Gutiérrez et al. (2015), and Cabrera et al. (2021), have together made a compelling case that multifractal cascades should be pervasive in neural and behavioral dynamics, leading to seemingly complex patterns at higher-levels of behavioral organization, such as spatial searching patterns observed in foraging animals including humans. Speaking of the empirical support, Suckling et al. (2008), for example, have shown that the multifractal spectrum of endogenous fMRI activation—which serves as a proxy for executive function—can refract the effects of task performance, age, and even cholinergic blockade using scopolamine on brain activity. In addition, La Rocca et al. (2018) have reported multiplicative cascades in the resting-state and task-related MEG data along with a fronto-occipital gradient reminiscent of the known hierarchy of temporal scales from sensory to higher-order cortices. They have also shown that the observed gradient of cascades is more pronounced during the task than in the resting state, suggesting that multifractal cascades support the flexible and adaptive poise necessary for healthy executive functioning. We would also direct the readers to Ghosh et al. (2019), who have reviewed a body of work showing how disease-induced changes in multifractal cascades affect executive functioning.

Multiplicative cascades have also been shown to support visuomotor dexterity under changing constraints. For example, multifractality moderates the relationship between temporal correlations and performance in **Trail Making Test (TMT)** (Kelty-Stephen et al., 2016). Anastas et al. (2014) have provided evidence of multiplicative interactions across timescales in the hand motions of preschool-age participants sorting cards based on rules that changed without explicit warning or explicit instruction. In adults, this multifractal structure of hand movements also changed across time as participants mastered the task, and these changes unfolded differently across time depending on the type of instruction (D. Stephen et al., 2012). In the same vein, Arsac (2021) studied the leg and hand motions of elite rugby players with high motor skills concurrently performing rhythmic motor tasks involving cycling with their legs and tapping or circling with their arms. They have reported that not only do these movements show multiplicative interactions across timescales, but measures of multifractal dynamics also remain consistent across single tasks of cycling, tapping, or circling alone, or dual tasks of cycling with tapping and cycling with circling. Multifractality thus indexes the multiplicative cascades supporting a poise allowing adaptation to more complex tasks. Finally, Pratviel et al. (2021) provide further evidence that eye-hand coordination in the task of hitting spatially aligned targets under time pressure arises from multiplicative interactions across timescales of activity within the whole perception-action system. Collectively, these findings support the call for describing the control of visuomotor dexterity in terms of multiplicative cascades dynamics and a system-wide distributed control instead of a central executive 'talking' to isolated, encapsulated components. This system-wide distributed control can be simply understood in terms of tradeoffs between overly regular and overly random dynamics, which confers upon the system the capacity to navigate a complex world in which surprising events can occur at all timescales (Ihlen and Vereijken, 2013, 2010; Van Orden, 2010; Van Orden et al., 2005).

## 5. Cascade instability in postural control: Bodily poise for turning plans into action

Cascade instability may underwrite the bodily poise that serves as a wellspring for converting our plans into action in the world around us. Postural control of upright standing can seem like mere maintenance of position, but truly, it holds within it clues for how to spring outwards and extend our bodies in to a task context. As before with Turing's pattern arising from fluctuations, we have what might sound like another poetic turn of absurdity: a metaphor of fluid tumblings-down might strike us initially as logically incompatible with the task of stabilizing our upright posture. However, a growing empirical work has analyzed measured **postural sway** or **postural center of pressure (CoP)**, and converged around two significant points. First, postural sway exhibits the kind of power-law scaling and heavy-tailed PDFs consistent with cascading processes and inconsistent with synthetic surrogate data mimicking the linear properties of the measurements. Second, the empirical estimates of cascade instability provide compelling predictors of postural outcomes, suggesting that cascade-driven models might inform an explanation of postural control. Hence, it is not that upright standing bodies are tumbling down a rock face, but instead, the fluctuations in postural sway propagate across many scales across the body, for instance, from the large tilt of the torso or turn of the head to the subtle twitch of muscle fibers or the adjustment of ankles and toes. In a sense, we can think of cascades as the forms that our body takes on to absorb and quickly release fluctuations that might destabilize posture. We can also envision these cascading bodily forms as a resource for resetting our posture to fit the task context. In much the same way Turing's patterns arose from instability, the cascading instability of posture is critical for producing new bodily patterns to suit the ongoing activity.

As suggested by the summary of empirical analysis of cascade structure, the focus of cascade-inspired research on postural control has been on the waxing and waning of postural fluctuations over wide ranges of timescales. This approach is a marked departure from the more traditional technique of assessing the amplitude of sway deviation. That is, cascade-inspired postural research asks the empirical question of "What shape?" with the expectation that the shape of sway will be hierarchically organized, for example, with finer details branching from or nested within larger details. And the traditional approach has largely only asked "How much?" with the expectation that sway is homogeneous enough to be well represented by a standard deviation. The cascade-inspired approach has proven helpful in demonstrating the various entailments of multiplicative cascades for various task constraints and individual characteristics such as disease state. For example, for participants fixating gaze on a visible target, the distance to the target modulates the multifractality of body sway during upright posture. Older people in this task exhibit more evidence of multifractality than younger adults, suggesting that cascading leads tiny fluctuations to propagate and persist longer through the postural system (Munafo et al., 2016). Other studies have have reported a similar distinction between obese and nonobese children, which is amplified by closing the eyes (Fink et al., 2016), and between healthy adults and patients with balance disorders, which is amplified by the removal of visual or tactile information (Shimizu et al., 2002).

The underlying cascade structure may speak empirically to how postural sway supports suprapostural dexterity in a task context (Kelty-Stephen et al., 2021). Estimates of cascade structure show that multiplicative interactions across timescales allow the body to assume the poise for suprapostural tasks flexibly. For instance, cascade instability supports upright posture in the service of maintaining visual fixation on a target in the visual field (Mangalam et al., 2021). It not only prevents larger fluctuations in sway but grows stronger as the target strays from the most comfortable seeing and reaching distance (Fig. 8). Hence, cascade instability provides glimpses of a close-knit relationship between the visual layout and the postural system, supporting our engagement with objects and events. This nested cascade-like structure is in some sense an ongoing response to the nested structure of optic flow in the visual field. In one study, for example, experimentally reducing optic-flow nesting with blinders blocking the visual periphery diminishes estimable cascade structure in postural sway during Fitts tasks with manual pointing or comfortable walking (Eddy and Kelty-Stephen, 2015). Then again, no such diminution appeared in participants whose sway explored beyond the blinders' bounds with more head sway that was more planar (e.g., circular) than ballistic (e.g., side-to-side or back-to-front). The availability of nesting focal vision within the periphery fostered nonlinear interactions across nested timescales so that endogenous bodily exploration can also foster and rebuild.

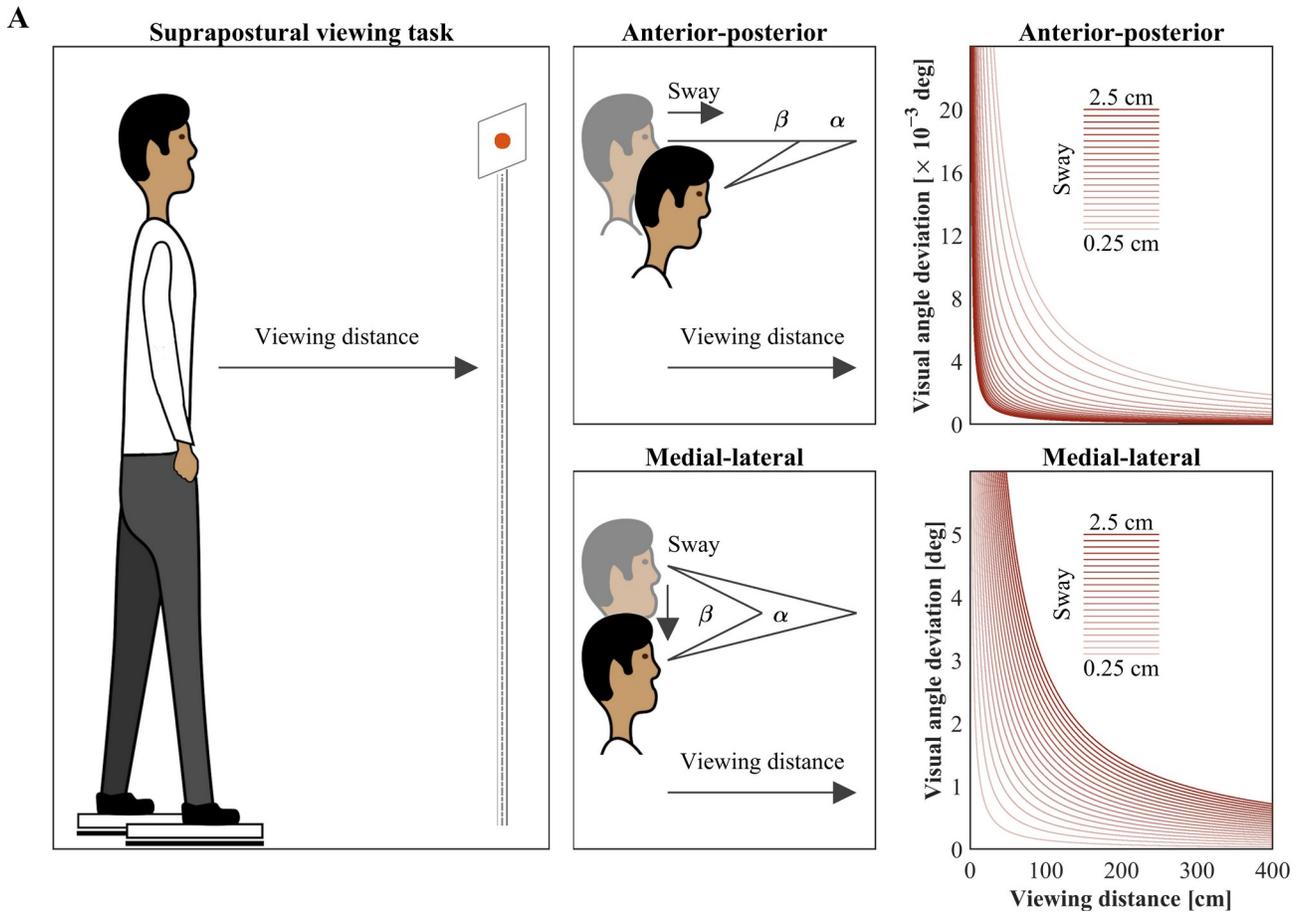

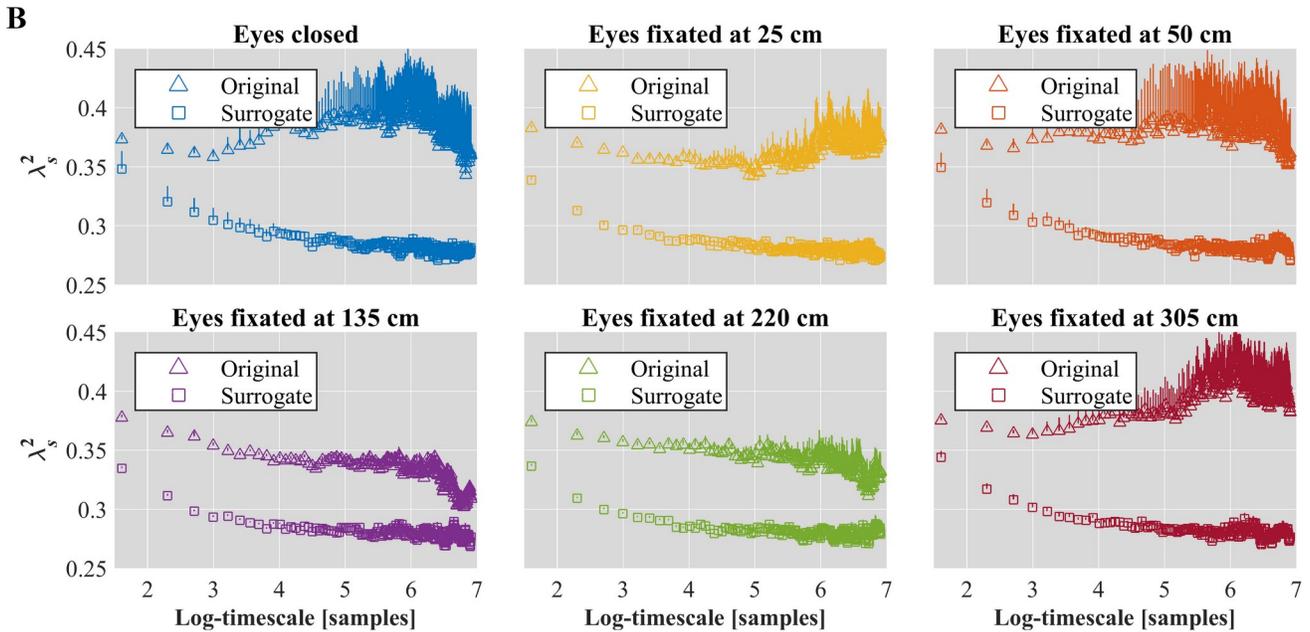

**Fig. 8.** Schematic of the task and effects of eye-to-target distance on postural sway as revealed by multiscale PDF analysis. (A) The suprapostural viewing task of standing quietly with the eyes fixated at a distant visual element (left); visual angle gain for short vs. long eye-to-target distances along the anterior-posterior (*AP*) and medial-lateral (*ML*) axes (middle); and visual angle gain as a function of eye-to-target distance for different sway magnitudes (right). Closer targets increase *AP* sway,

whereas farther targets increase *ML* sway. (**B**) Log-timescale dependence of the non-Gaussianity index $\lambda_s^2$. Mean values of $\lambda_s^2$ in postural CoP for the participants standing quietly for 120 s in different viewing conditions: Eyes closed; Eyes fixated at 25 cm; Eyes fixated at 50 cm; Eyes fixated at 135 cm; Eyes fixated at 220 cm; Eyes fixated at 305 cm. Vertical bars indicate ±1*SEM* ($N$ = 15 participants). The $\lambda_s^2$ vs. log-timescale curves differ between the eye-closed condition all the eyes-open conditions except the eyes-fixated-at-50-cm condition—that is, comfortable viewing distance elicited posture with a similar cascade dynamics as posture with eyes closed. Adapted from Mangalam et al. (2021).

This cascade-driven implication of nested, interacting timescales offers the possibility that postural poise for action leaps forward from a foundation that reaches out to multiple timescales at once. That is, the point of postural contact with the layout for action is not simply the foot's current, momentary plantar surface contact with the ground or the shoe insole. On the contrary, the cascade structure of posture entails that each momentary contact of the postural system with the context is a nexus of several ongoing processes. The apparent instability or dexterity of postural may thus result from the fact that many of these ongoing processes are complicating, mutually supporting, or competing. However, the cascade structure of the postural away extends the postural system beyond the current moment and spatial coordinates and enfolds the posture into the broader layout of the future. Therefore, it is not surprising that multiplicative-cascade dynamics play a role in suprapostural prospective coordination, appearing in the fluctuations in using the fingertip to balance pole-balancing (Harrison et al., 2014) or attempts to finger presses with mathematically unpredictable events (Kelty-Stephen and Dixon, 2014).

Although the cascade is by definition a broadly constructed unfolding across multiple timescales, we can be pretty specific about the physiological forms supporting their appearance and governing their progression. Cascades are not material-specific; no single tissue alone can embody a cascade. Instead, just as a neurotransmitter operates by having the right shape to fit into a post-synaptic neuron, the physiology here exerts its control through morphology, that is, through geometry. The multiplicative cascades in postural sway and their entailment in these stabilization tasks are deeply rooted in the **tensegrity**-like structures (Turvey and Fonseca, 2014). 'Tensegrity' is a portmanteau term linking "tension" with "integrity," denoting a prestressed construction of the body that embodies nonlinear interactions across spatial and temporal scales (Fig. 9; see also https://www.youtube.com/watch?v=eW0lvOVKDxE&t=454s). As noted above, the recipe for the cascade embodied by a tensegrity is not material but formal: tensegrities compromise a balancing of tension and compression elements across many different scales, and often, the compression elements at one scale serve as tensional elements at another (Ingber, 2006). Tensegrity systems encompass genetic and neural dynamics and include a vast array of physiological processes.

They encompass a rich network of connective tissues that support prospective control of posture even in systems with minimal nervous systems. Furthermore, cascade-like multiplicative interactions across timescales support postural control even in insect models with sparse nervous systems but quick prestressed responsivity (Kelty-Stephen, 2018).

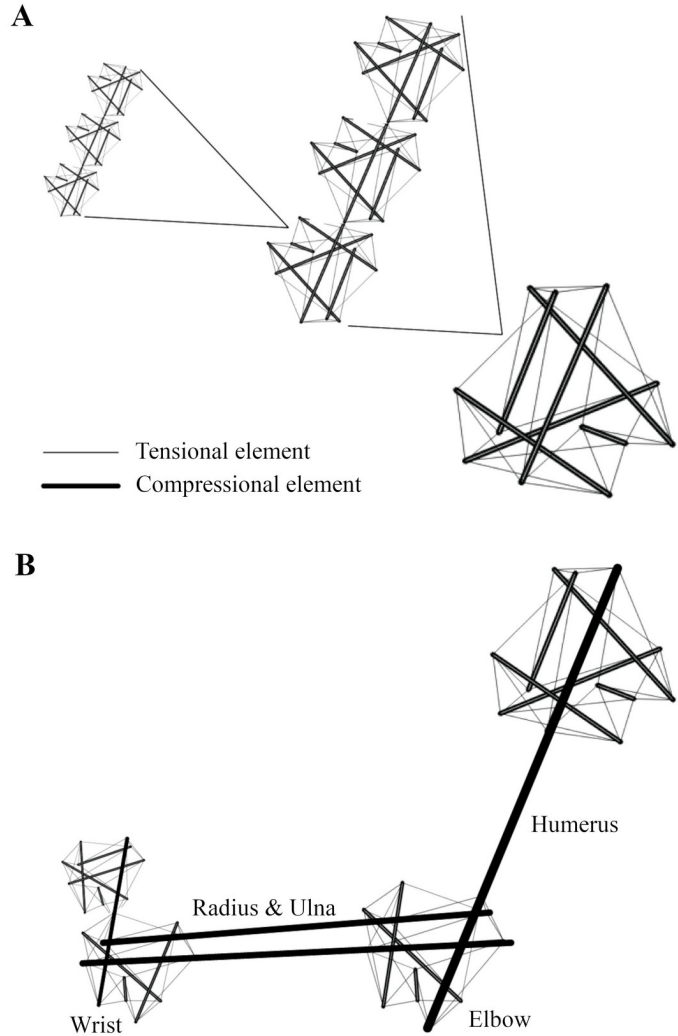

**Fig. 9.** Tensegrity view of the body. (**A**) A tensegrity icosahedron (left), and a multifractal tensegrity (MFT) icosahedron with multifractality spanning across scales (middle & right). (**B**) A tensegrity model of the human arm. Any local perturbation stimulates a global response. The core notion is that under tensional and compressional stresses, the force is applied or transferred across all elements of the system, allowing shape, solidity, multidirectional movement, and gravity independence. Tensegrity structures are supported not by the resistance of each element, such as a column resisting gravity, but by the distribution and equilibrium of mechanical stresses across the entire network. As a result, such structures are multidirectional, stable in all directions, and gravity-independent. See Turvey and Fonseca (2014) for the MFT hypothesis to body-brain linkages and explanations of various neuropathological consequences.

Besides addressing the specific physiology supporting a cascade, we can also elaborate the cascade model to be more functionally specific. For instance, we can identify the specific ranges of timescales over which postural cascades unfold using multiscale PDF analysis. This approach improves upon prior dominant theorizing in postural control that had sometimes envisioned sway as reflecting two separate modes of control: ballistic, inertial open-loop vs. feedback-dependent closed-loop control (Collins et al., 1995; Collins and De Luca, 1993). The precision of cascade-based analysis like multiscale PDF allows us to see that, on the contrary, destabilizing posture does not entail a digital switch between modes of control. Instead, destabilizing posture entails a flexible reorientation with cascade-like sensory corrections migrating fluidly between the longer scales of so-called closed-loop and the shorter scales of so-called open-loop control. This capacity for cascades to extend across some or all of the timescales appears similarly in measured postural sway no matter whether we load the upper body (Furmanek et al., 2020) or destabilize contact with the ground surface (Mangalam and Kelty-Stephen, 2021).

Another refinement of a cascade-driven model of postural control appears in the feature of scale-in/dependence. Scale-in/dependence is not a dichotomous feature: cascades can show gradually different degrees of independence from the scale. Cascades can produce scale-invariant forms showing similar distributions and interactions across many scales (e.g., Jensen, 1998; Van Orden et al., 2003). However, cascade flow bends and buckles under various constraints (e.g., choke points, valleys, or boundaries) that exert their effect at specific scales (Lovejoy, 2019; Pattee, 2007). Perturbing a cascade at a specific scale can make the cascade scale-dependent, and we can confirm this point in empirical measurements of postural control. Of course, we can take a relatively coarse multifractal analysis, compare multifractal results between original and surrogate data, and determine coarse qualities like whether or not cascade-like instability plays a nonzero role (Kelty-Stephen and Mirman, 2013). However, multiscale PDF analysis allows us to conceptualize the overall cascade patterning and resolve deeper, finer structures to inform theoretical models. Cascades comprise multiple streams that can align with separable postural-control functions like "**rambling**" and "**trembling**." Rambling is the relatively global capacity to reset equilibrium points across the ground surface, and trembling reflects the control local to maintaining a given equilibrium point. Cascade modeling of posture shows a concomitant matching of global and local functions (Fig. 10). Constraints on posture accentuate the strength of scale-dependent cascade processes in trembling aspects of posture while leaving scale-invariant cascade processes unchanged in rambling aspects of posture (Kelty-Stephen et al., 2020).

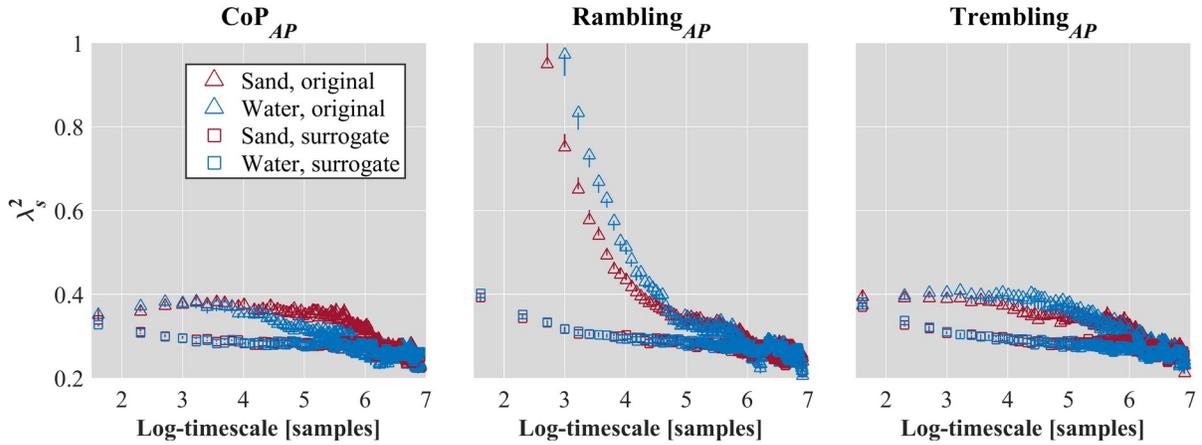

**Fig. 10.** Log-timescale dependence of mean $\lambda_s^2$ for the postural CoP, rambling, and trembling trajectories along the anterior-posterior (*AP*) axis in the postural tasks of standing quietly while balancing sand- and water-filled tubes for 30 s. Vertical bars indicate ±1*SEM* (*N* = 10 participants). Adapted from Kelty-Stephen et al (2020).

The disparate analytical methods available to model cascade instability appear to converge on a unifying portrait of postural control. That is, we do not simply have multiple separate empirical tools that can each find their isolated sign of a cascade. Furthermore, we are not simply in a position of, for instance, desperately consulting another analysis if the first comes up with no evidence of cascades. On the contrary, and encouragingly, the capacity of multiscale PDF analysis to distinguish varying degrees of scale-dependence goes hand in hand with the multifractal analysis already developed to inquire into cascade dynamics. Specifically, Kelty-Stephen et al. (2020) found that multifractal spectrum width of postural-control measurements moderated the waxing and waning of scale-dependence in both trembling and rambling cascades. Estimates of the multifractal structure live up to the long-offered promises that might help reveal how cascades operate. The cascade-instability formalism has grown a set of tools that can peer deeply into our behavioral-science measurements, and we can speak to long-standing questions in the theoretical discourse around them. So, as noted above, we can confirm that cascade instability supports a 'rambling vs. trembling' spectrum within postural control even while it undermines any 'open vs. closed-loop' distinction. Then again, we can bring new substance to the postural-control theory, as with the insight that cascade-driven multifractality might be a common substrate supporting the differentiation of postural control into rambling and trembling subsystems.

## 6. Cascade instability in effortful perception: Action to gather information, in a single modality

As bodies step forth from a quiet stance and reach farther than sway allows, organisms clamber about their surroundings. The turbulent, creative tension between possibility and constraints unfolds, with organisms balancing their ability with their intentions, aiming to explore the constraints for opportunities and obstacles alike. Furthermore, the instabilities that unfold reveal yet more of the same cascades we had seen in posture. Cascade dynamics characterize how bodies explore their surroundings to arrive at perceptual judgments and, more generally, develop perceptual relationships with events in the world. We might encounter this "effort" most explicitly in the domain of "effortful" or "dynamic touch" using muscular stretch to sense various properties of wielded objects (e.g., heaviness, length, width, shape, and extent along different dimensions) . However, we can also elaborate this notion of effort supporting perception to a wider variety of model systems enlisting effort for orienting postures and movements towards the environmental structure. Throughout, fractal and multifractal evidence promises to reveal how cascade instability spanning bodily degrees of freedom supports the detection, pooling, and sharing of perceptual information within and between organisms. Hence, cascade dynamics offers a compelling way to begin envisioning embodied, situated models of perception founded in whole organisms or groups of organisms.

The journey to a full-body view of perception can begin with a single point of bodily contact with stimulus information. We can begin this approach by asking the simplifying question of what mechanical information means when explored only by a single movement-system effector. If cascade instability evokes a turbulent tempest, then dynamic-touch paradigms, in effect, make an elegant effort to corral this tempest into the metaphorical teapot where we can begin to examine under experimental constraint. Operationally, this simplification comes from occluding curtain aligned with the wielding arm's shoulder, allowing experimenters to separate—perhaps artificially but pragmatically—the haptic awareness from visual awareness. We can, for present purposes, manipulate the inertial properties of the objects held in hand, and the occluder gives us a theoretical room in which to discuss the role of distinctly mechanical properties separate from visual properties. We can thus begin to reduce haptic perception to a mechanical contact local to one part of the body. We can, in effect, attempt to focus the whole-organism drama of exploration and information into a single movement-system effector, grasping and wielding an unseen object behind a curtain (Turvey and Carello, 2011).

Swept behind the curtain, the cascade-born tempest of effort in perception can seem no more daunting than the writhings and wrigglings of a hand, clutching the object handle and swinging it this way and that way. Following a deep psychophysical tradition, we can examine the intrinsic features of the object and model power-function relationships between the objects' kinetic properties (e.g., rotational inertia) and perceptual impressions (e.g., perceived length; Fitzpatrick et al., 1994). However, this work leaps beyond traditional psychophysics in its foundational aims: to anchor such stimulus-perception relationships in the fluid, self-organizing physics of an intentional organism (Solomon and Turvey, 1988). Under slightly different intentions, the organism can mine the same physical events involving the grasping, wielding hand for different information specific to different facts of the environmental layout (Carello et al., 1992). Moreover, different modes of exploration might reveal different kinds of information (Kingma et al., 2004) and indeed, a deeper point may lie in the evidence that intention will change the exploratory movements (Arzamarski et al., 2010). So, in a sense, try as we might fix the parameters of the mechanical information through the tailoring of objects, the organism betrays a fundamental fluidity. On the other side of the curtain, visual feedback about size judgments prompts organisms to tailor their exploratory movements to retune their use of informational variables to better align judgments with feedback (Michaels et al., 2008; Wagman et al., 2008). The organism commands a degree of flexibility that no experimental teapot can restrain.

At this point, as we watch the writhing and wriggling hand behind the curtain, we may ask new questions about what may not be plain to our eyes—is all wielding the same if it can leverage different classes of information and support different responses? Have the "apparently selfsame physical events" offering different information been somehow different—even in some nonobvious way the experimenter's eye may not catch? Indeed, the calculations of object properties involve the Euclidean approximations, for example, to points where object mass is balanced and points in the wrist that anchor presumed rotational axes. These approximations are effective for homogeneous, solid stimulus objects, but they fit inelegantly on human movement systems whose survival and dexterity depend on fluidity and heterogeneity. The coarse approximation has long held sway over-attention to the fluid error terms in such approximations. The leading scholars in dynamic touch research have long been champions of developmental theories like that of Gilbert Gottlieb (2007, 1998, 1997) emphasizing the role of nonobvious factors percolating through a perceptual system (Turvey and Fitzpatrick, 1993; Turvey and Sheya, 2017; Wagman, 2010). Curiously, however, the attention to nonobvious aspects of muscular exertion in this domain has been limited, for example, to speculation about the role and form of tremor in static holding (Burton and Turvey, 1990) and isolated portraits of the time series properties of wielding behavior (Riley et al., 2002).

However, a closer look at the limbs grasping and wielding an object suggests that cascades pervade effortful perception no less than they pervade postural control. Try as we might distill the exploration into a single point of contact with mechanical information, the exploration betrays the same cascade instability as we saw in the upright posture. The wielding movements are fractally scaled, with the displacements between consecutive hand positions showing correlations across all measured timescales. Wielding is not simply a patchwork of independent wiggles and twitches. Instead, wielding exhibits a temporal structure that, on each trial, spans the whole bout of exploration leading up to the judgment, and the subtle movements at more acceptable timescales show a similar patterning of fluctuations as we find over the longer timescales (Stephen et al., 2010).

More importantly, the cascade instability of these exploratory behaviors predicts how participants use the same mechanical information to arrive at perceptual judgments. And the fractal structure is not simply an exotic by-product but may reveal cascading roots of perception: trial-by-trial variation in the fractal structure of wielding predicts trial-by-trial variation in perceptual judgments. These fractal variations help to predict how perceptual judgments change across varied task settings, with judgment type (e.g., length or width; Stephen et al., 2010), by exploratory limb (e.g., hand or foot; Stephen and Hajnal, 2011), with attention deficits (Avelar et al., 2019), and with visual feedback about judgments (Kelty-Stephen and Dixon, 2014; Stephen et al., 2010). A unifying theme of this work is that the multiplicity of fractal profiles, that is, the "multifractality" in this motoric cascade, allows us to predict how participants make use of available information. Information might start "out there," but just as bodily scaling frames, whether available surfaces afford stepping or sitting (Mark, 1987; Warren, 1984; Warren and Whang, 1987), multifractality of exploration exemplifies how diversity in bodily geometry supports diverse perceptual impressions.

Cascade instability does not just persist across various task settings but spreads across the whole body—try as we might contain the cascading tempest of exploration in the experimental teapot of an occluding curtain, the whole organism gets into the cascading exploratory act. Indeed, the occluding curtain offered to simplify our explanatory path by examining only a single point of entry for stimulation. Then again, organisms are nothing if not more than the sum of independent anatomical parts (Kelty-Stephen et al., 2021; Lewontin, 1982; Turvey and Fonseca, 2014). Reducing our view of haptic perception to the local point of stimulation is a reasonable simplification. True, the occluding curtain allows distinguishing between visual and haptic information, but cascades brewing in exploratory wielding hand movements stem from—and ripple back across—the rest of the organism. The bodily response to an object held by the hand is thoroughly global. Occluding the grasped object does not decompose the organism into separable perceptual subsystems (e.g., visual and haptic). Instead, it perturbs a bodywide cascade coursing across disparate motoric degrees of freedom. Furthermore, perhaps disturbingly, when participants use hand wielding to perceive various properties of unseen objects while standing on a force plate, the multifractality in the postural center of pressure (CoP) predicts their perceptual judgments (Mangalam et al., 2020b; Mangalam and Kelty-Stephen, 2020; Fig. 11). Undoubtedly, the object is in hand and not underfoot, but what can the feet' relationship with the ground surface do with effortful touch by the hand? Certainly, for a participant standing perfectly still, the multifractality of CoP has direct implications for perceiving objects supported by the shoulders (Palatinus et al., 2013) and differs as they attend to all or part of the object supported by the shoulder (Palatinus et al., 2014). Nonetheless, even though it can seem intriguing to find so much multifractal texture in "standing still," the contact looks local: shoulders are centered over the CoP. What do we make of the connectivity between the grasping hand and standing feet? Is this just a downstream echo, with CoP reverberating later with what the wielding hand does before?

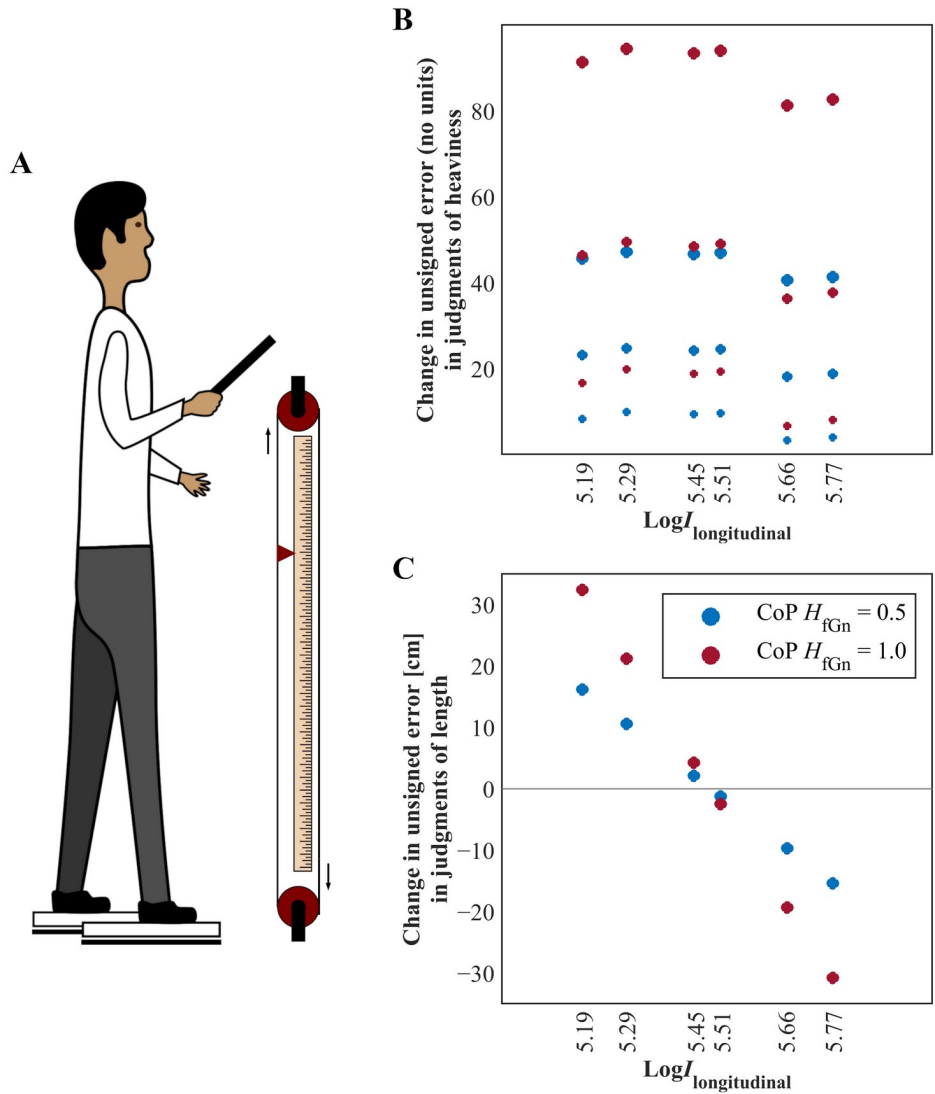

**Fig. 11.** The strength of fractal fluctuations—estimates using Hurst exponent $H_{fGn}$—in the Euclidean displacement in the postural CoP contributed to perceptual judgments by moderating how the participant picked up the informational variable of the moment of inertia specifying that object's heaviness and length. (**A**) Each participant stood with their feet on separate force plates, wielded each object for 5 s, and reported their judgements of heaviness of that object by speaking out a number (e.g., a judgement of 50 or of 200 would indicate an object half or twice as heavy as the reference object, respectively) and length by adjusting the position of a marker on a string–pulley assembly. (**B**, **C**) Changes in unsigned error in judgments of heaviness and length for each object. The sizes of solid circles in (B) indicate progressively increasing levels of the torque. Each point includes an error bar indicating *SD* of the unsigned errors for each corresponding object ($N$ = 15 participants), but each bar is nearly invisible because *SD* of the unsigned error was much smaller within object than across objects. See Mangalam et al. (2020b) for further details.

The flows of multifractal fluctuations at different points of the body reveal how the movement system pools the information drawn from different sources.

Multifractality in hand wielding is rooted in multifractal fluctuations in muscular activity. For instance, multifractality of muscular activity in the wielding arm is predictive of perceived length—more predictive of those perceptual judgments than just the amount of muscular activity (Mangalam et al., 2019b, 2019a). In addition, on the allegedly non-wielding side of the occluding curtain, participants' heads sway as they look at the response apparatus, negotiating with pulley-and-string to position the marker correctly. Though entirely uninstructed and minimal enough to go unnoticed, this head sway has a multifractal patterning that may be crucial for making use of visual feedback. In fact, visual feedback may show no noticeable effect except when we include estimates of this multifractality in the statistical modeling (Kelty-Stephen and Dixon, 2014). Of course, experimenters may balk here and protest that the head sway is beyond the instructed intention when instructed to wield by the hand. However, if we could entertain the head sway as wielding, we might see new ways to experimentally elaborate the long-held notion of upright postural sway as generating optic flow (Riccio and Stoffregen, 1991; Stoffregen and Riccio, 1990). The body and head may each wield an optic array, not unlike the haptic array wielded by the hand. Intriguingly, multifractality in head sway alone can mediate how participants translate between apparently disparate senses, such as visual or haptic experience to haptic or visual response, respectively (Hajnal et al., 2018). For instance, multifractality in head sway supports judgments about visually or haptically available slopes, from affordance judgments to confidence judgments of those affordances, to the use of affordance and confidence judgments to inform continuous-scaled estimates of slopes (Doyon et al., 2019).

## 7. Cascade instability in effortful perception: Acting to blend multimodal information

Blending haptic with visual information to make a visible response about mechanically available events depends on organisms fostering specifically different multifractality at different body parts. The organism is not a monolith, and the variations of multifractality across different body parts are vital for allowing vision (e.g., through head sway) to have a different impact than haptics (e.g., through wielding by the hand). These collaborative relationships between multifractality in the head and hand hold in less contrived cases that do not involve an experimenter's occluding curtain, for example, visually aiming to a distant target (Carver et al., 2017) and the Fitts' task asking participants to make alternating finger contact with visible targets (Bell et al., 2019). Moreover, in both cases, we find torso-sway multifractality playing a similar role as head sway in mediating between visual and mechanical information (Eddy and Kelty-Stephen, 2015; Jacobson et al., 2021). Torso-sway multifractality also supports the nonvisual perception of distance by walking, i.e., when the target is distant but not seen (Teng et al., 2016). There are indeed many points of activity across the body participating in the coordination of multiple ways to grapple with the available information.

Coordination and sharing of information might rely on spreading multifractal fluctuations from one anatomical part of the body to another to relay information—or, at least, consequences of perceptual exploration. Some of these relationships offer novel generic support for known relationships among apparently different modalities. For example, for participants wielding objects they cannot see, increases in head-sway multifractality promote multifractality in later wielding by the hand; this relationship may be key to lacing visual into haptic information. Experimentally introducing visual feedback leads head-sway on hand-wielding multifractality, prolonging the later increase in hand-sway multifractality much more than participants without feedback (Kelty-Stephen and Dixon, 2014). Similarly, experimentally perturbing visual information with prismatic right-shifting goggles during visually-aimed throwing behaviors requires throwing hands to draw upon the novel visual-and-haptic orientation of the body. It is thus intriguing to note that throwing-hand multifractality increases at the expense of later multifractality at the hip and head. We can also consider effortful perception at other parts of a developmental trajectory. A commonplace assumption is that infants are just learning to use their movement systems and use early movement variability to explore their limbs and the surrounding information (Berger et al., 2019; Rademacher et al., 2008). Multifractal modeling has shown that spontaneous kicking movements in infants show an upstream flow of multifractal fluctuations from the ankle to knee and so on to the hip (Stephen et al., 2012).

Cascade instability thus brings a view of effortful perception as the accomplishment of a bodywide swarm of activity. Indeed, one organism has one body, but the bodily degrees of freedom may constitute a veritable swarm of quasi-autonomous parts capable of falling in and out of cooperation. This almost-social view of within-organism coordination could sound too much like the communication among separate organisms, but this analogy could be a risk worth taking. For instance, overtly language-based behaviors appear to thrive on auditory sensitivity to multifractal structure in speech sounds (Bloomfield et al., 2021; Wallot et al., 2015; Ward and Kelty-Stephen, 2018) and on the multifractality of movements used to explore linguistic stimuli (Bloomfield et al., 2021; Booth et al., 2018; Wallot et al., 2015). Moreover, for non-human swarms, multifractality of group behavior can index success or failure to coordinate (Dixon and Kelty-Stephen, 2012), and multifractality of individual-organism behavior can predict group membership better than linear features of behavior (Carver and Kelty-Stephen, 2017).

Taking this swarm perspective on the single body may resolve our earlier question of what postural sway could have to do with wielding by the hand. Specifically, it can reveal that postural CoP multifractality is not merely an echo of multifractality at hand. Instead, postural CoP multifractality could be a potential source of multifractality at hand, above and beyond various other multifractal relationships across the body. We can consider motion-capture data from a marker set widely distributed across the body, including joints across both hands and arms and locations across the shoulders and torso. If effortful perception were just about the hand exploring an object, all of these non-hand marker positions might recover no useful information. However, what we had only considered above in pairs or triples of head, hand, and torso now unfolds across our view of the whole-body movement (Mangalam et al., 2020a). The wielded object and wielding arm both course with multifractal fluctuations, and each appears to foster later increased multifractality in each other. Multifractal fluctuations appear to flow both down the forearm to the object and also from the object back to the upper arm. And then it appears that, above and beyond all of these relationships across the upper body, increases in multifractality of postural CoP predict later increases in hand wielding (Fig. 12).

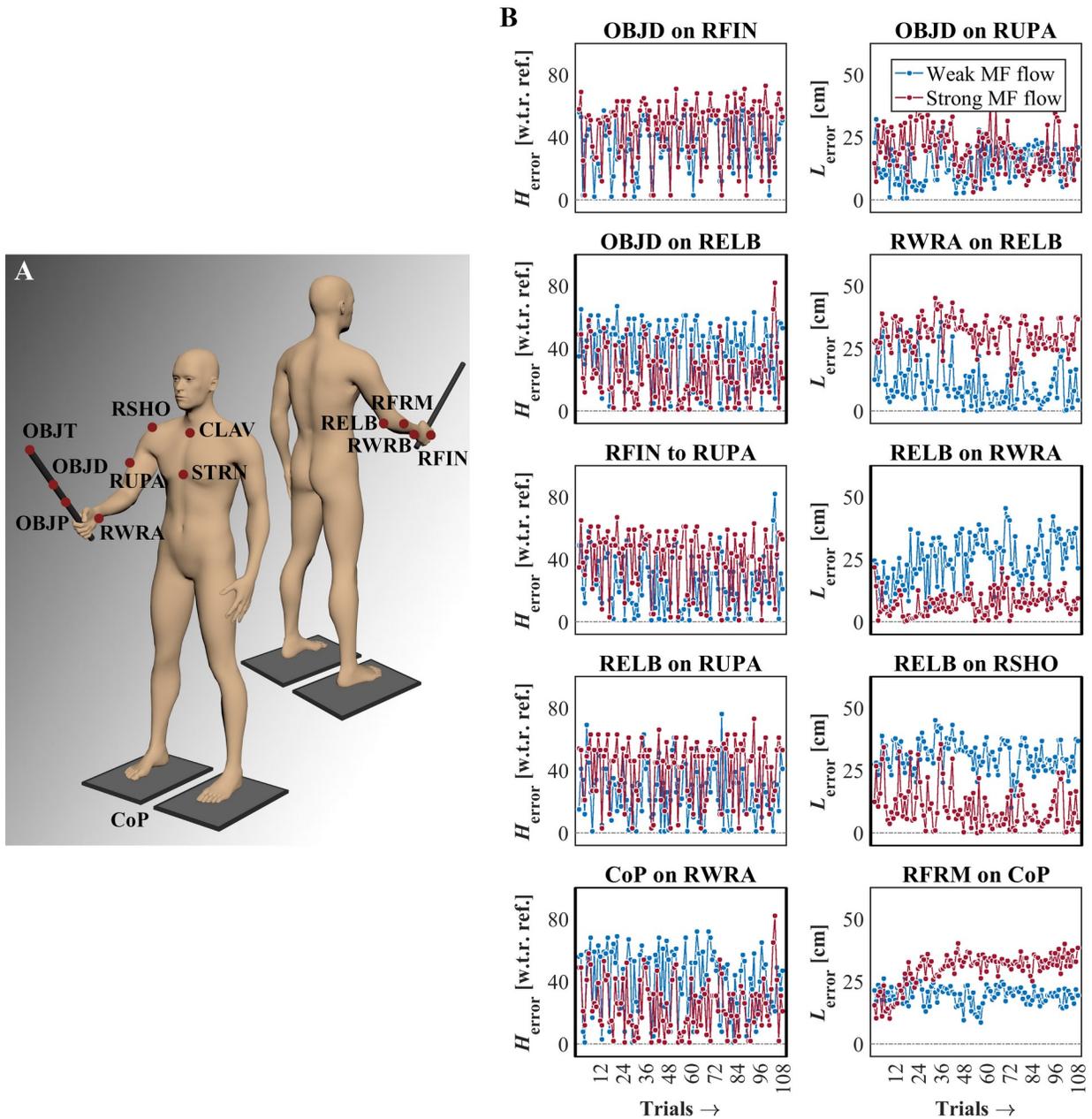

**Fig. 12.** The flow of multifractal fluctuations across the body supports perception of object heaviness and length by dynamic touch. (**A**) Each participant stood with his/her two feet on separate force plates, wielded each object for 5 s, and reported his/herjudgements of heaviness of that object by speaking out a number (e.g. a judgement of 50 or of 200 would indicate an object half or twice as heavy as the referenceobject, respectively) and length by adjusting the position of a marker on a string–pulley assembly. Sample-to-sample fluctuations in movement of several body parts and the postural CoP were recorded and submitted to multifractal analysis followed by causal network analysis. (**B**) Comparisons of absolute errors in perceived heaviness, $H_{error}$, and perceived length, $L_{error}$, for representative participants with weak and strong flow of multifractal fluctuations from one body part to another. The strong IRF effects of OBJD on RELB and CoP on RWRA entailed decrease in $H_{error}$ (left panels in bold); all other IRF effects entailed

increases in $H_{error}$ (left panels). The strong IRF effects of RELB on RWRA and RELB on RSHO entailed decrease in $L_{error}$ (right panels in bold); all other IRF effects entailed increases in $L_{error}$ (right panels). See Mangalam et al. (2020b) for details.

Cascade instability thus serves to knit the whole body together, from plans for action to bodily poise to gathering information relevant to the task at hand. It offers a fluid framework in which individual parts of the same body can link together or flow apart to serve a perceptual goal. Cascade instability affords the organism, at once, the unitary integrity and the internal heterogeneity that has so long challenged attempts to formalize organisms as a distinct class (Baedke, 2019; Baedke et al., 2021; Lewontin, 1982; Wilson, 2000). Cascade instability affords a capacity to diversify its information search and then coordinate these diverse types of information, making the seemingly multisensory faculties a unitary perceptual system (Stoffregen et al., 2017). Multifractal measures of cascade instability begin to play their predictive role by indexing the richness of exploration by any single point of contact between organism and task context. Nonetheless, on continued observation, the cascade instability then begins to spill over the rim of our experimental teapots, thwarting our well-planned constraints. Events at many isolated degrees of freedom will generally find a way to return to the bodywide swarm that hums with activity spanning the multiple contacts and investments organisms make with and in their context. Cascades may be a sufficiently generic mechanism to explain the congealing and coordination of multiple degrees of freedom, from plan to poise for action to information gathering.

## 8. Neuronal avalanches in the brain: Neural participation in bodywide cascades

The portrayal of bodily dynamics in the language of cascade instability from plan to action to perceptual judgment raises the critical question of what sort of control system would such processes obey? The cascading bodily dynamics leave room for and would benefit most from a nervous system capable of participating in these cascades—perhaps with cascades of its own. Neuroscientists have long sought to articulate how the nervous system could best serve the swarm-like blurriness between plurality and integrity of organism behavior. They have entertained speculation that the nervous system may thus benefit from organization principles that put it at what is called a "critical point," a regime between order and disorder.

Much as the flow of cascade instability respects the anatomical connectivity across the body, neuronal avalanches unfold according to the geometry or architecture of the neural network (Fig. 13). Criical-state dynamics of avalanches arise in specific neural networks embodying a specific balance between excitation and inhibition (Gireesh and Plenz, 2008; Mazzoni et al., 2007; Pasquale et al., 2008; Seshadri et al., 2018; Stewart and Plenz, 2006). Neuronal avalanche dynamics rely on functional and structural small-world topology and its corresponding traffic in cortical networks (Pajevic and Plenz, 2009), reflecting sensitivity to structural heterogeneity (Benayoun et al., 2010; Kello, 2013; Levina et al., 2007; Lombardi et al., 2017; Millman et al., 2010; Scarpetta and de Candia, 2013; Wu et al., 2019). Homeostatic plasticity mechanisms in healthy brain networks are now thought to control this heterogeneity at many levels of neuronal architecture (Poil et al., 2012, 2008), possibly via transient cell assemblies (Plenz and Thiagarajan, 2007).

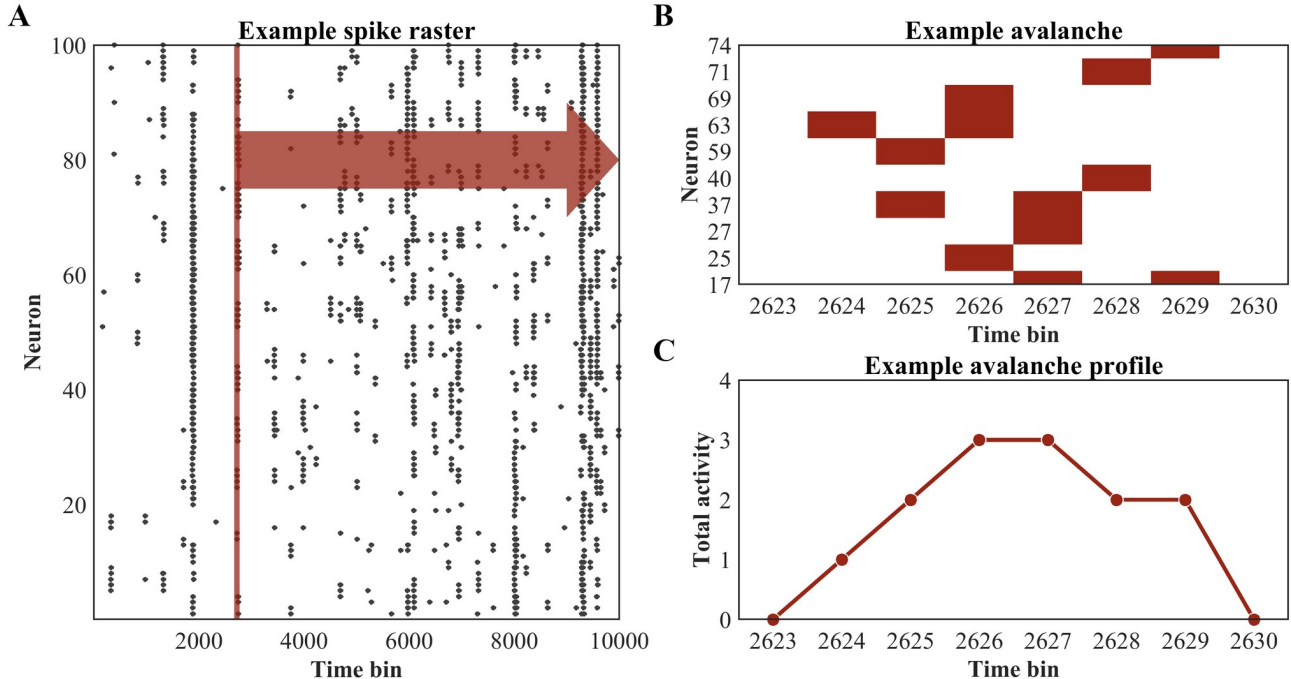

**Fig. 13.** Neural avalenches. (**A**) A segment of the spike raster for all neurons in an example cortical branching model. (**B**) Example neuronal avalanche. Adjacent periods of activity were identified as avalanches. This avalanche corresponds to the red vertical line in (**A**) and is 6 time bins long and involves 13 total neuron activations. (**C**) Avalanche profile for the avalanche shown in (**A**). Adapted from Marshall et al. (2016).

The alignment between cascade instability and neuronal avalanches persists in a shared reliance on the same multifractal geometry, that is, the appearance and often diversity of power-law distributions. The null hypothesis for the neuronal avalanche literature is a **Poisson-like process** in which an ensemble of independent, uncoupled neuronal units would yield exponential distributions of event sizes. The appearance of power laws is the predicted outcome of multiplicative interactions in a distributed network's collective activities (Alstott et al., 2014; Hartley et al., 2014; Klaus et al., 2011; Ribeiro et al., 2014; Taylor et al., 2013; Timme et al., 2016; Touboul and Destexhe, 2017, 2010). However, the focus on the existence or not of power laws has given way to more nuanced ways to understand the growth or change of these power laws. For instance, recent research has expanded the number of analysis techniques (Beggs and Timme, 2012; Marshall et al., 2016) to include **shape collapse** (Friedman et al., 2012; Priesemann et al., 2009), **susceptibility** (Gleeson and Durrett, 2017; Williams-García et al., 2014), and **neural tuning** through the critical point (Shew et al., 2011, 2009). More recently, there has been a greater appreciation of the fully multifractal form of neuronal avalanches (Alamian et al., 2022; Bogdan, 2019; Bonachela et al., 2010; Das and Levina, 2019; Gilpin, 2021; Jannesari et al., 2020; Miller et al., 2019; Papo et al., 2017; Zorick et al., 2020), as well as persistent calls to examine and acknowledge a more generic class of cascade processes to address more diverse outcomes (Bonachela et al., 2010; Gilpin, 2021; Papo et al., 2017).

The neuroscience literature has also begun to develop neuronal cascades into modeling tools that can reach back out into the behavioral tasks that we saw suffused with cascades above. In a rare but revelatory move, Gutierrez and Cabrera (2015) explicitly identified an alignment between the multifractal cascades of neuronal activity and body-wide cascades generating fractal patterns of navigating a task. The fractal patterns of fMRI BOLD signals likewise can predict variation in executive function tasks (Kardan et al., 2020b, 2020a). In both cases, we see the beginnings of a connection from cascades in the brain to cascades in the other bodily or performance domains. The connection is incomplete because these results stop short of the complete portrayal of multifractality in the brain or of multifractality in behaviors, but we hope that this alignment through the common substrate of cascade instability signifies future possibilities for novel insights into the feats of adaptive, context-sensitive organisms.

## 9. Concluding remarks

A full empirical accounting of the value of cascade instabilities for the mind, brain, and behavior remains forthcoming. Nonetheless, we hope to have made the case that the instances of cascade instability offer the potential of a unifying formalism reaching across the many tissues and functions of the perceiving, acting, and thinking organism. In a sense, there is nothing new here but a reaffirmation of Turing's old insights. And it may also be thoroughly unsurprising and likely not coincidental that bodily tissues and neuronal activity should bear geometrically similar signatures across various functions and outcomes. But given the potential obviousness of the analogy among cascades in planning behaviors, in poise, perceptual information gathering, and control, we do find it alarming that the potential role of cascade instability in shaping the relationship between body and brain remains relatively unexplored, except for speculation (Turvey and Fonseca, 2014) and only a handful of empirical demonstrations (Gutiérrez and Cabrera, 2015; Kardan et al., 2020b, 2020a).

We suspect that some of the challenges and possibilities ahead lie in another of Turing's insights—also old news but maybe holding lessons we are only learning. Namely, we find ourselves sitting between the complementary modes of logical and physical science. There is a habit for us to think of Turing's work as split, for example, between "the two sides of Turing" (Kelso, 1995, p. 4). The logical, rate-independent processes characterize the computational process that never tires, never deviates, and never starts or stops except at the command of the computer user. And the physical, rate-dependent processes progress by exhausting finite resources and can respond to or change course with the flimsiest perturbation or buffeting breeze (Hodges, 1983). The symbol-like constraints of logical processes have suffered a second-class status in many physical descriptions, possibly because what looks like a constraint at one scale may be reduced to a physical process at another (Pattee, 2001). For anyone who has decided which scales might matter and which might be left to themselves, for example, as in Simon's (2019) proposal of near-decomposability, there may be no reason to concern ourselves with what this or that piece looks like to an observer. If we have chosen our preferred vantage point on the system, then the logic and the physics may as well sit separately.

However, it is a mistake to see these logical and physical aspects as separate. They were, after all, part of the same visionary organism's work, and much though posterity has understood these strands as separate, Turing (1950) entertained the possibility that these complementary threads were mutually supporting parts of the same weave. The utility of symbols and logic requires physics to support the processing of the logical constraints and the interpretation of the logical constraints. This latter point implicates other scales in a way that leaves room for skepticism about near-decomposability: the smaller-scale electrons support the processes of digits through a computer, and the larger-scale movements we make (e.g., keypresses, scrolls, and browsings) support how we take any meaning from the computer. The symbols are somewhere in between, whether as the polarity of electrons or as the 1-ness or 0-ness in a binary sequence. These symbols are effectively logical: as Pattee (2001) suggested, they are rate-independent because they will be there when I return to the screen after an interruption but then only become useful in complementing one another (Pattee, 2013).

The cascade instability is a physical mechanize Turing (1950) proposed to grow novel logic. If there are any doubts about whether Turing meant cascades, Turing specifically invoked "criticality" as the source of novel logic. If there were analytical and mathematical obstacles to modeling this instability and its creative powers, then Turing (1952) and subsequent work examining interactions across scales supporting growth and "criticality" have paved a way forward. For instance, we know now that the interactions across scales that Turing himself (Dawes, 2016) invoked yield fractal and multifractal patternings (Mandelbrot, 1974). Multifractal geometries for exploring power laws, their variety, and their emergence or disappearance have arrived and are doing empirical work where they are taken seriously (Lovejoy and Schertzer, 2018). In this sense, the path is clear toward asking the empirical questions about whether the cascade instability Turing pioneered organizes the whole circus of mind, brain, and behavior.

Any remaining obstacles to progress in this direction do not look empirical; they may sooner be theoretical or metaphorical. It is worth considering whether our metaphors for the brain have constrained this progress. Have we perhaps taken the computer metaphors for the brain too literally? Have we done so with too deep a trust in near-decomposability? Indeed, we see that there are theories of development—even psychological development—that rely on a turbulent interweaving of multiple scales of activity (Gottlieb, 2007; Turvey and Sheya, 2017).

Some of these theories even strike a resonant chord with the earliest and most poetic metaphors for the mind in modern psychology. For instance, William James's idea of the mind as a river may carry some new force. Lyrical and opaque as it might have been back in James's day, it remains with us today—perhaps in new terminology, such as a "mountain stream" (Thelen and Smith, 2006). We hope that the cascades that have long been recognized as characteristic of the mind—the fluidity, the interactivity, the capacity to generate structure across scales and not simply within them—may give birth to a relatively new way to understand Turing's cascade instability and its capacity to bring logical and physical into the explanatory complement.

**Funding**

We received no funding for this work.

**Competing interests**

We declare we have no competing interests.